\documentclass[preprint,sort&compress,3p]{elsarticle}

\usepackage{url}
\usepackage{amsmath}

\usepackage{graphicx}
\usepackage{multirow}
\usepackage{multicol}
\usepackage{float}
\usepackage{subfigure}
\usepackage{hyperref}

\usepackage{fancyhdr}
\renewcommand{\thispagestyle}[1]{} 

\newcommand{\numu}[0]{$\nu_\mu$}
\newcommand{\nutau}[0]{$\nu_\tau$}
\newcommand{\nue}[0]{$\nu_e$}

\newcommand{\antinumu}[0]{$\overline{\nu}_\mu$}
\newcommand{\antinutau}[0]{$\overline{\nu}_\tau$}
\newcommand{\antinue}[0]{$\overline{\nu}_e$}

\newcommand{\thetaatm}[0]{$\theta_{23}$}
\newcommand{\dmatm}[0]{$\Delta m^2_{32}$}
\newcommand{\thetasol}[0]{$\theta_{12}$}
\newcommand{\dmsol}[0]{$\Delta m^2_{21}$}
\newcommand{\dmreac}[0]{$\Delta m^2_{31}$}
\newcommand{\thetareac}[0]{$\theta_{13}$}
\newcommand{\deltacp}[0]{$\delta_{CP}$}

\newcommand{\ket}[1] {\ensuremath{| #1 \rangle}}

\begin{document}

\pagestyle{fancy}  
\lhead{}  
\rhead{}  

\title{Neutrino Oscillations with MINOS and MINOS+}
\author[lhw]{L. H. Whitehead\fnref{collab}}
\ead{l.whitehead@ucl.ac.uk}
\fntext[collab]{For the MINOS collaboration}
\address[lhw]{Department of Physics and Astronomy, University College London, Gower Street, London, WC1E  6BT, United Kingdom}

\begin{abstract}
The MINOS experiment ran from 2003 until 2012 and collected a data sample including 10.71$\times10^{20}\,$ protons-on-target (POT) of beam neutrinos, 3.36$\times10^{20}\,$POT of beam antineutrinos and an atmospheric neutrino exposure of 37.88$\,$kt-yrs. The final measurement of the atmospheric neutrino oscillation parameters, \dmatm{} and \thetaatm{}, came from a full three flavour oscillation analysis of the combined CC \numu{} and CC \antinumu{} beam and atmospheric samples and the CC \nue{} and CC \antinue{} appearance samples. This analysis yielded the most precise measurement of the atmospheric mass splitting \dmatm{} performed to date. The results are  $|$\dmatm{}$|=[2.28 - 2.46]\times10^{-3}\,$eV$^{2}$ (68\%) and $\sin^{2}$\thetaatm{}${}=0.35-0.65$ (90\%) in the normal hierarchy, and $|$\dmatm{}$|=[2.32 - 2.53]\times10^{-3}\,$eV$^{2}$ (68\%) and $\sin^{2}$\thetaatm{}${}=0.34-0.67$ (90\%) in the inverted hierarchy. The successor to MINOS in the NO$\nu$A era at FNAL, MINOS+, is now collecting data mostly in the $3-10\,$GeV region, and an analysis of \numu{} disappearance using the first 2.99$\times10^{20}\,$POT of data produced results very consistent with those from MINOS. Future data will further test the standard neutrino oscillation paradigm and allow for improved searches for exotic phenomena including sterile neutrinos, large extra dimensions and non-standard interactions. 
\end{abstract}

\begin{keyword}
Neutrino Oscillation\sep Long-baseline\sep MINOS\sep MINOS+
\end{keyword}

\maketitle

\section{Introduction}

Nearly two decades have passed since the first observation of neutrino oscillations by Super-Kamiokande~\cite{superKPRL1998}. In that time it has become very clear from a number of experiments looking at neutrinos from the sun, the atmosphere, nuclear reactors and man-made neutrino beams that neutrinos can undergo oscillations from one flavour to another~\cite{superKPRL2004,superKPRD2010,sno,k2k,minosPRL2006,minosPRD2008,t2kPRL2014NuE,t2kPRL2014,kamland,borexino,dayaBay2012,reno2012,doubleChooz2012}, as described by the PMNS matrix~\cite{mns1962,ponte1968,ponte1969}. The PMNS matrix, $U$, commonly parametrised by three mixing angles (\thetaatm, \thetasol{} and \thetareac) and a \emph{CP}-violating phase (\deltacp{}), describes the mixing between the three weak flavour eigenstates, $\ket{\nu_\alpha} $, and mass eigenstates, \ket{\nu_i} in the following way:

\begin{equation}\label{eq:flavMass}
\ket{\nu_\alpha} = \sum_{i=1}^3 U_{\alpha i} \ket{\nu_i}.
\end{equation}

The three mixing angles have been measured to varying degrees of accuracy but the value of \deltacp{} is still unknown. The oscillations arise from the quantum mechanical interference between the neutrino mass states and are driven by the mass-squared splittings between these mass states, $\Delta m^{2}_{ij}\equiv m^{2}_{i} - m^{2}_{j}$. It is possible to write down three mass-squared splittings, but only two are actually independent. One of the mass splittings, \dmsol{}, is considerably smaller than the others, meaning there are two scales at which oscillations can occur. The signs of the other mass-splittings, \dmatm{} and \dmreac{}, are currently unknown, meaning it is not known whether $m_3$ is the lightest or heaviest mass state. The case where it is the heaviest (lightest) is referred to as the normal (inverted) hierarchy. A final, important consequence of neutrino oscillations is the requirement that at least two of the neutrino mass states must be non-zero.

The two main oscillation channels of interest in long-baseline neutrino oscillation experiments are \numu$\rightarrow$\numu{} disappearance and \numu$\rightarrow$\nue{} appearance. These channels were first probed using a man-made neutrino beam by the K2K experiment~\cite{k2k,k2kNuE2004}. The discovery of \numu$\rightarrow$\nue{} oscillations was performed by T2K~\cite{t2kPRL2014NuE} and \numu$\rightarrow$\nutau{} appearance was discovered by the OPERA experiment~\cite{operaPRL2015}. Oscillations in such experiments are driven by the two larger mass-splittings, \dmatm{} and \dmreac{}. Using a two neutrino approximation, with the parameters $\Delta m^{2}$ and $\sin^{2}2\theta$, the \numu{} disappearance probability for a neutrino with energy E and travelling over a distance L in the vacuum can be written as follows:

\begin{equation}\label{eq:2flavDis}
P(\nu_\mu\rightarrow\nu_\mu) = 1 - \sin^{2}2\theta\sin^{2}\left(\frac{\Delta m^{2} L}{4E}\right).
\end{equation}

However, \thetareac{} was measured by Daya Bay~\cite{dayaBay2012} and later by RENO~\cite{reno2012} and Double CHOOZ~\cite{doubleChooz2012} and is hence known to be reasonably large. In addition, the uncertainty on measurements of \dmatm{} is of the same order as the size of \dmsol{}. The more accurate formalism requires the use of the full three flavour oscillation probabilities and the approximate parameters $\Delta m^{2}$ and $\sin^{2}2\theta$ in Eq. \ref{eq:2flavDis} are modified in the following way~\cite{3flavForm}:

\begin{align}\label{eq:3flavDis}
  \sin^{2}2\theta &= 4\cos^{2}\theta_{13}\sin^{2}\theta_{23}(1-\cos^{2}\theta_{13}\sin^{2}\theta_{23}), \nonumber \\
  \Delta m^{2} = \Delta m^{2}_{32} + &\sin^{2}\theta_{12}\Delta m^{2}_{21} + \cos\delta_{CP}\sin\theta_{13}\sin2\theta_{12}\tan\theta_{23}\Delta m^{2}_{21}.
\end{align}

The expressions given in Eq. \ref{eq:3flavDis} illustrate how the interference between the two different mass-splitting terms causes the full oscillation probability to depend on all of the parameters of the PMNS matrix. It can be seen in Eq. \ref{eq:2flavDis} that the two flavour oscillation probability is symmetric under the transformations of $\theta \rightarrow \frac{\pi}{2} - \theta$ and $\Delta m^{2} \rightarrow -\Delta m^{2}$. The equivalent parameter shifts for the three flavour case in Eq. \ref{eq:3flavDis} are $\theta_{23} \rightarrow \frac{\pi}{2} - \theta_{23}$ and $\Delta m^{2}_{32} \rightarrow -\Delta m^{2}_{32}$, and it can be seen that the oscillation probability is not completely symmetric under these transformations, leading to approximate degeneracies instead of symmetries.

When neutrinos traverse matter, the Hamiltonian associated with the propagation is modified compared to that of the vacuum by interactions of the neutrinos with the matter. All three neutrino flavours can undergo neutral-current (NC) interactions with the matter via the exchange of a $Z$ boson but since this affects all neutrinos equally, it does not cause a change in the oscillations. However, only electron neutrinos can have charged-current (CC) interactions with the electrons in the matter via the exchange of a $W$ boson, giving rise to a change in the oscillations. This phenomenon is known as the MSW effect~\cite{wolf1978,ms1986}. In this case, $\theta_{13}$ is replaced by a modified mixing angle $\theta_M$ as defined by~\cite{matterShift}

\begin{equation}\label{eq:matterEffect}
\sin^{2}2\theta_{M}=\frac{\sin^{2}2\theta_{13}}{\sin^{2}2\theta_{13} + (\cos2\theta_{13}-A)^2},
\end{equation}
where $A = 2\sqrt{2}G_{F}n_{e}E/\Delta m^{2}_{31}$, $G_{F}$ is the Fermi weak coupling constant and $n_{e}$ is the electron density. In the case of antineutrinos, the value of $A$ changes from $A\rightarrow-A$. It is clear to see that when $\cos2\theta_{13} = A$, the value of $\sin^{2}2\theta_{M}$ becomes maximal, producing a resonance in the oscillation probability for \numu{}$\rightarrow$\nue{} oscillations, and hence modifies the \numu{} disappearance probability as well as the \nue{} appearance probability. This resonance occurs in multi-GeV atmospheric neutrino events that travel upwards through the earth's mantle, and since $A$ is dependent on the sign of $\Delta m^{2}_{31}$, it provides a handle with which to study the neutrino mass hierarchy.

The \numu{}$\rightarrow$\nue{} oscillation probability in matter, calculated up to second order in $\alpha = \Delta m^{2}_{21}/\Delta m^{2}_{32}$ is given by the following expression~\cite{nueForm}:

\begin{align}\label{eq:3flavApp}
P(\nu_\mu\rightarrow\nu_e) \approx \sin^{2}\theta_{23}\sin^{2}&2\theta_{13} \frac{\sin^{2}\Delta(1-A)}{(1-A)^2} +\alpha\tilde{J}\cos(\Delta\pm\delta_{CP})\frac{\sin\Delta A}{A}\frac{\sin\Delta(1-A)}{(1-A)} \nonumber \\
&+\alpha^2\cos^{2}\theta_{23}\sin^{2}2\theta_{12}\frac{\sin^{2}\Delta A}{A^{2}},
\end{align}
with $\tilde{J}=\cos\theta_{13}\sin2\theta_{13}\sin2\theta_{12}\sin2\theta_{23}$ and $\Delta = \Delta m^2 _{31} L/4E$. The positive sign of \deltacp{} in the second term refers to neutrinos, whilst the negative sign corresponds to antineutrinos. Equation \ref{eq:3flavApp} shows that the \numu{}$\rightarrow$\nue{} appearance channel is sensitive to: the octant of $\theta_{23}$ through the first term, the \emph{CP}-violating phase through the presence of \deltacp{} in the second term, and the mass hierarchy from the matter effect parameter, $A$.

\section{The MINOS/MINOS+ Experiment}
The Main Injector Neutrino Oscillation Search (MINOS) experiment was originally designed in order to accurately measure the atmospheric parameters of neutrino oscillations, namely \thetaatm{} and \dmatm{}. MINOS began collecting atmospheric neutrino data in 2003 and beam data-taking began in 2005. The experiment ran until May 2012 when the beam was shut off in order to prepare for the NO$\nu$A experiment. At this point, MINOS transitioned into MINOS+, the name of the experiment going into the NO$\nu$A era. MINOS+ began collecting beam data in September 2013 when the beam switched back on.
\subsection{The NuMI Beam}
The Neutrinos at the Main Injector (NuMI) beam~\cite{numiBeamPaper} is the neutrino beam at the Fermi National Accelerator Laboratory (FNAL) that supplied neutrinos for MINOS, and currently produces neutrinos for NO$\nu$A, MINER$\nu$A and MINOS+.

\begin{figure}
\centering
\includegraphics[scale=0.35]{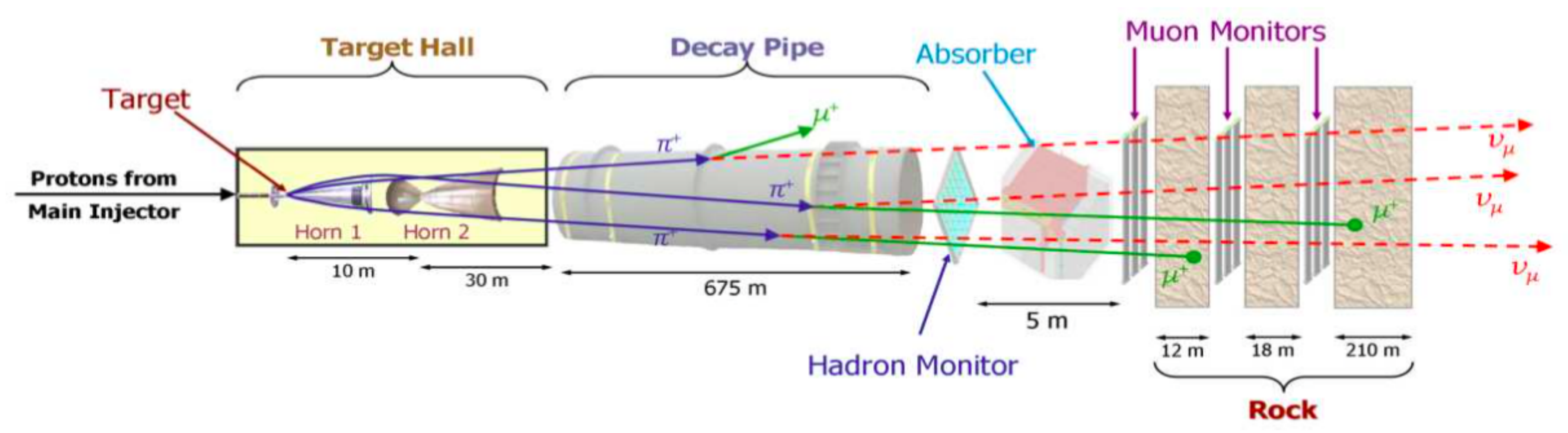}
\caption{\label{fig:numi}Schematic diagram of the NuMI beam line showing the major components with their corresponding size (not to scale). The beam line is actually oriented downwards into the ground at an angle of 58$\,$mrad to the horizontal in order to point towards the Far Detector. The Near Detector cavern is downstream of the last section of rock on the right side of the figure.}
\end{figure}

The main components of the NuMI beam are shown in Fig. \ref{fig:numi}. Protons with an energy of 120$\,$GeV are extracted from the Main Injector (MI) proton accelerator and are steered onto a graphite target. The spray of hadrons, primarily pions and kaons, resultant from the collisions of the protons with the carbon nuclei are focussed by two current-pulsed magnetic horns and directed into the decay pipe.

The focussed hadrons then decay at some point along the 675$\,$m decay pipe to produce the neutrino beam. Muon neutrinos are mostly produced by the following decays:

\begin{equation}\label{eq:piDecay}
\pi^{+} \rightarrow \mu^{+} + \nu_\mu
\end{equation}
\begin{equation}
K^{+} \rightarrow \mu^{+} + \nu_\mu
\end{equation}
and electron neutrinos by the decays of kaons and tertiary muons:

\begin{equation}\label{eq:muDecay}
\mu^{+} \rightarrow e^{+} + \nu_e + \overline{\nu}_\mu
\end{equation}
\begin{equation}
K^{+} \rightarrow e^{+} + \nu_e + \pi^{0}
\end{equation}
\begin{equation}
K^0_L \rightarrow e^{+} + \nu_e + \pi^{-}.
\end{equation}

The charge conjugate processes also exist to produce the antineutrinos but are suppressed by the focussing of positively charged mesons in nominal neutrino beam mode. 

Downstream of the decay pipe is the hadron monitor that measures the spatial distribution of any remaining hadrons. The absorber, formed from an aluminium core, with a steel and concrete surround, is located downstream of the hadron monitor and stops any remaining hadrons (mostly protons from the beam that did not interact and some mesons that did not decay in the decay pipe). It is only the muons and neutrinos that pass through the absorber, and the muons are then measured with three muon monitors interspersed with regions of the natural dolomite rock. The total 240$\,$m of rock upstream of the Near Detector (ND) cavern stops all of the muons, leaving a beam consisting only of neutrinos and antineutrinos.

It is possible to change the position of the target and the magnetic horns in order to change the energy distribution of the beam. The vast majority of the MINOS data were taken in the Low Energy (LE) beam configuration where the target was partially inserted into the first magnetic horn, giving a neutrino beam peaked at approximately 3$\,$GeV. 

In standard neutrino mode operation the magnetic horns are set such that they focus positively charged mesons, resulting in a neutrino beam and known as Forward Horn Current (FHC) running. It is possible to reverse the current used to pulse the horns in order to focus the negatively charged mesons to produce a beam with an enhanced antineutrino component, a configuration known as Reverse Horn Current (RHC). In FHC (or \numu{}-dominated) mode, the beam consists of 91.7\% \numu{}, 7.0\% \antinumu, and 1.3\% \nue{} $+$ \antinue{} and in RHC (or \antinumu{}-enhanced) mode, 58.1\% \numu{}, 39.9\% \antinumu, and 2.0\% \nue{} $+$ \antinue{}~\cite{minosApp2012}. Short periods of data were taken in other configurations in order to study the beam. 

\begin{figure}
\centering
\includegraphics[scale=0.45]{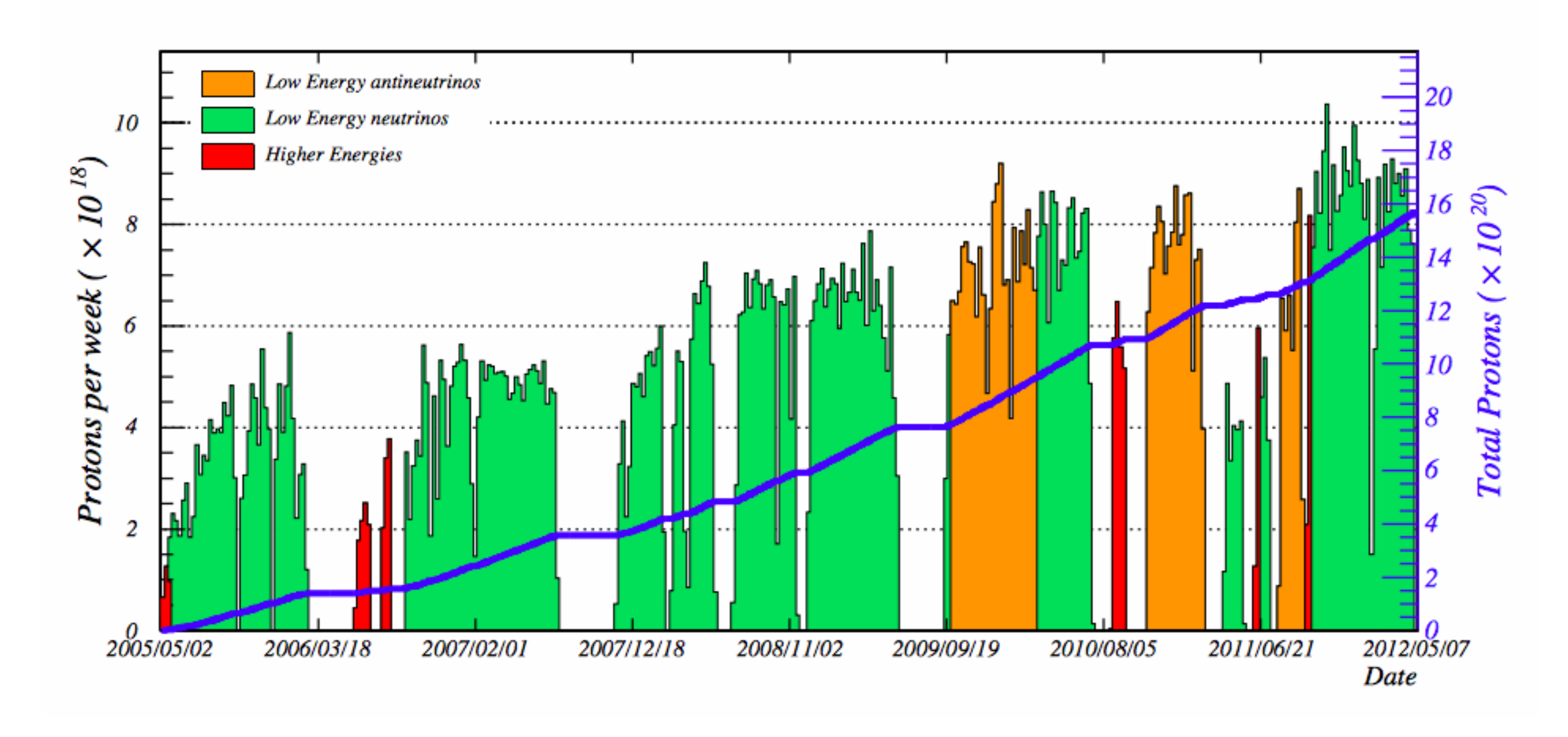}
\caption{\label{fig:numiPOT}The total accumulated POT (blue line) and the number of protons per week (histograms) delivered by the NuMI beam from May 2005 until May 2012. The green regions show the POT delivered in standard LE mode in the FHC configuration. The orange shows the antineutrino running, and the red shows special short runs in different configurations. Figure from Ref.~\cite{numiBeamPaper}.}
\end{figure}

The NuMI beam supplied a total of 10.71$\times10^{20}$ protons-on-target (POT) in FHC mode and 3.36$\times10^{20}$ POT in RHC mode to the MINOS experiment. Figure \ref{fig:numiPOT} shows the number of protons delivered per week and the total accumulated POT as a function of time from May 2005 until May 2012. The POT in FHC mode is shown in green, and the orange shows the data in RHC mode. Short special runs, such as those with the magnetic horns turned off or at higher energy, are shown in red. As of September 2013, the NuMI beam is operated in Medium Energy (ME) mode to supply neutrinos to NO$\nu$A, an experiment that is off-axis from the beam, and MINOS+. The ME beam has a peak at about 6$\,$GeV on-axis for MINOS+. 

\subsection{Beam Flux Simulation}\label{sec:beamFluxSim}
\begin{figure}
\centering
\includegraphics[scale=0.38]{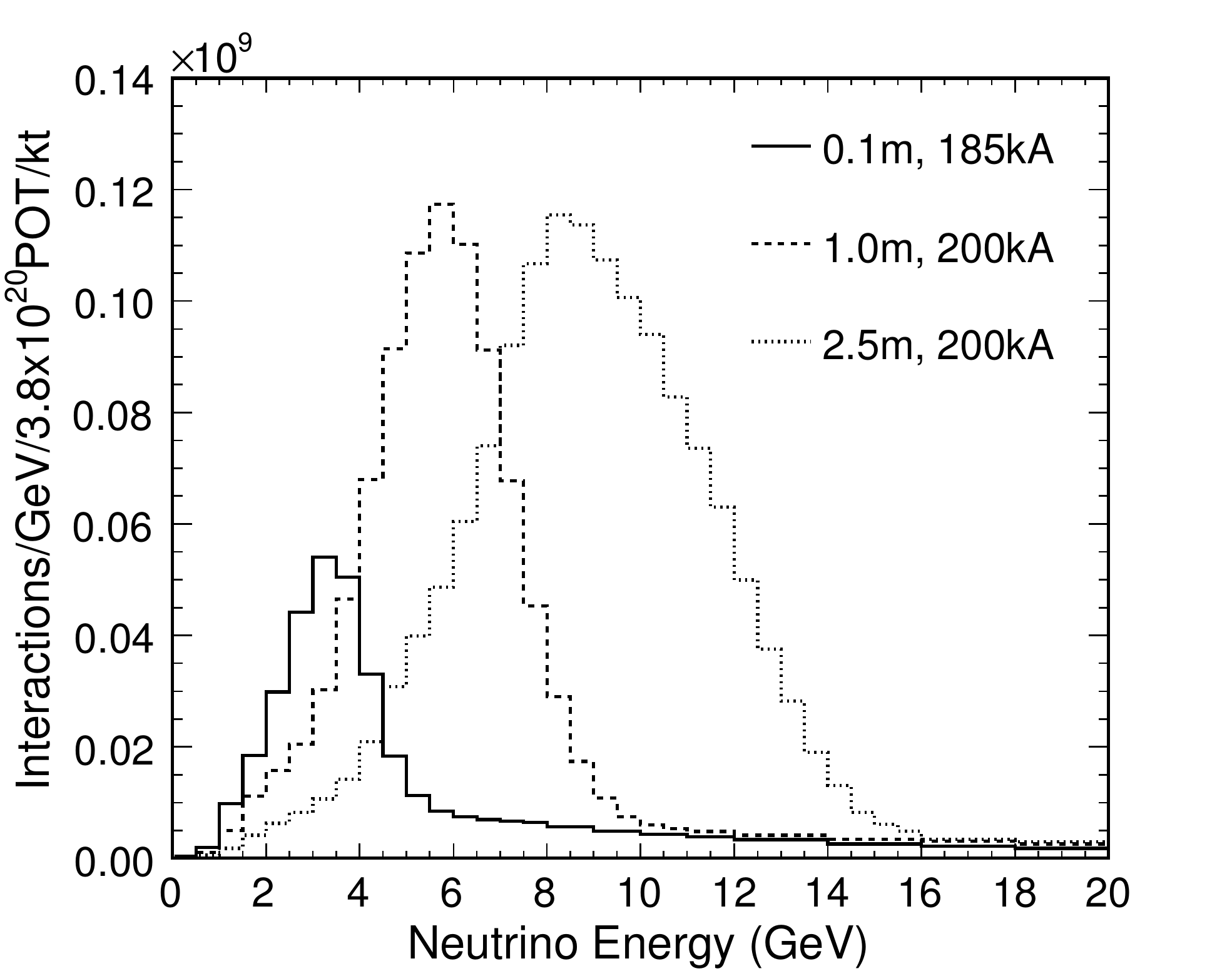}
\caption{\label{fig:fluxConfig}The true energy distribution of neutrino interactions in the ND from the simulation of three beam configurations. The solid line corresponds to LE running, the mode in which most MINOS data were taken. The dashed line shows the ME spectrum, as used in MINOS+, and the dotted line corresponds to pHE. Figure from Ref.~\cite{ashleyReview}.}
\end{figure}

The neutrino beam flux is simulated using a combination of the GEANT4~\cite{geant4} geometry package and the FLUKA~\cite{flugFluka} hadron production package known as FLUGG~\cite{flugFluka}. Figure \ref{fig:fluxConfig} shows the true energy distribution for simulated events in the ND for the LE (solid), ME (dashed) and pseudo high energy(pHE) (dotted) beam configurations. The beam simulation does not, however, provide a perfect description of the neutrino flux that is measured at the ND. As such, the ND is used to constrain the simulation, since there are underlying uncertainties, particularly in the hadron production in the target, that cause disagreement between data and simulation. This method is described in detail in Ref.~\cite{minosPRD2008}, but is outlined below.

Each bin of the reconstructed CC \numu{} energy spectrum in the ND contains neutrinos coming from the decay of meson parents that had given values of the transverse $(p_t)$ and longitudinal $(p_z)$ momentum components as they left the target. The different $(p_t,p_z)$ bins contribute to different energy bins, meaning that the ND data can be used to constrain the hadron production in the target. Special data samples were taken with different beam configurations to enhance the range of the $(p_t,p_z)$ space covered by the ND data to provide more accurate fits. This tuning procedure allows for the calculation of weights that are applied to the simulation, providing a more accurate description of the data.

\subsection{The MINOS Detectors}
MINOS has two functionally equivalent detectors~\cite{minosNIM} called the Near Detector and the Far Detector (FD). The ND, located at FNAL, lies 1.04$\,$km from the target and has a mass of about 1$\,$kt. The FD has a mass of 5.4$\,$kt, is located 705$\,$m (2070$\,$m water-equivalent) underground in the Soudan Mine, Minnesota, at a distance of 735$\,$km from the target.

The two detectors are magnetised steel/scintillator sampling calorimeters. They are formed from interleaved planes of 2.54$\,$cm steel to provide the interaction target mass and 1$\,$cm plastic scintillator to provide the active region of the detectors. The plastic scintillator planes are formed from bars that are 1$\,$cm $\times$ 4.1$\,$cm in cross-section and vary in length from $2.5\,$m to $8.0\,$m, and are read out via a wavelength shifting optical fibre that is embedded into the surface of the scintillator bars. The wavelength shifting fibres are read out using multi-anode PMTs, and are instrumented on one (both) ends for the ND (FD). The orientation of the scintillator bars on alternating planes are at $45^\circ$ and $-45^\circ$ to the vertical, to provide two orthogonal views, which along with the depth into the detector along the beam direction, provide 3D tracking information.

The magnetic field in each of the detectors is toroidal and provided by a current-carrying coil that passes through the middle of the detectors. The current direction can be reversed in order to change the charge of the particles that are bent inwards to the centre of the detector. The direction of the current is chosen to match the current in the magnetic horns such that negatively charged muons are focussed when running in FHC mode, and positively charged muons are focussed in RHC mode. The particles focussed into the detector generally have better energy resolution since range can be used to measure the momentum of contained particles, and for those that exit the detector, the longer path-length enables a better determination of the energy by curvature. The average magnetic field strength in the ND and FD is 1.3$\,$T and 1.4$\,$T, respectively~\cite{minosNIM}.

\subsection{Neutrino Interactions in the Detectors}\label{sec:minosInt}
There are three main types of neutrino interactions expected in the MINOS detectors:
\begin{itemize}
 \item CC \numu{} and \antinumu{}: The muon neutrino interacts with a nucleus $X$ via the exchange of a $W$ boson in the process $\nu_\mu + X \rightarrow \mu^{-} + X'$. These events are characterised by the track-like energy deposits caused by the muon, in addition to a hadronic shower at the interaction vertex. The separation of CC \numu{} and CC \antinumu{} interactions is performed by using the curvature to measure the sign of the muon charge.
 \item NC $\nu$: A neutrino scatters off a nucleus $X$ via $Z$ boson exchange $\nu + X \rightarrow \nu + X'$. Neutral current interactions appear purely as a hadronic shower, for all three neutrino flavours, since the scattered neutrino is not detected. With no charged lepton resultant from the interaction, it is not possible to distinguish between NC events involving different neutrino flavours.
 \item CC \nue{} and \antinue{}: The electron neutrino interacts with a nucleus $X$ via $W$ boson exchange in the process $\nu_e + X \rightarrow e^{-} + X'$.  These events appear as a small electromagnetic shower, and since the electron does not have a track-like topology, no separation between CC \nue{} and \antinue{} interactions can be made.
\end{itemize}
While a small number of CC \nutau{} events occur in the FD at high energy, it is very difficult to distinguish them from the event types listed above, such that no event selection is attempted.

\section{Muon Neutrino Disappearance}
MINOS can measure the atmospheric oscillation parameters \dmatm{} and \thetaatm{} by looking for the disappearance of muon neutrinos. MINOS is sensitive to CC \numu{} and CC \antinumu{} interactions from two sources: the NuMI beam and atmospheric neutrinos. This section outlines the methods used to select samples of these events from the two different sources.

\subsection{Beam Neutrinos}\label{sec:beamDis}
The method employed by the beam neutrino analysis is to use the ND to predict the expected FD reconstructed neutrino energy distribution for a given set of oscillation parameters in order to find the best fit values of the parameters. Beam muon neutrinos are selected in the ND and FD by looking for the track-like signature of the muon in charged-current \numu{} interactions. The complete sample consists of selections of the following types of events:
\begin{itemize}
 \item CC \numu{} interactions in the FHC beam.
 \item CC \antinumu{} interactions in the FHC beam.
 \item CC \numu{} rock and anti-fiducial muons (RAF) in the FHC beam.
 \item CC \antinumu{} interactions in the RHC beam.
\end{itemize}
All samples apart from the RAF selection require that the interaction vertex lies within the fiducial volume of the detector. The RAF selection aims to select those neutrino-induced muons that traverse the detector from neutrino interactions in the rock upstream of, and surrounding, the detector cavern, as well as those interactions that occur outside of the fiducial volume, close to the edge, of the detector. As such, RAF events consist only of a muon track that enters the detector from the outside, or a CC \numu{} interaction that occurred inside, but very close to the edge of the detector. In either case, only the muon is considered~\cite{rafThesis}.

Firstly, candidate events are considered only if they are in time with the beam and contain a track-like energy deposit. Four variables that describe the topology and energy deposition of the track are used as inputs to a k-Nearest-Neighbour (kNN) algorithm that produces as output a value between 0 and 1 that acts as the particle identification (PID) variable~\cite{pidThesis}. This PID is used to preferentially select the CC \numu{} and CC \antinumu{} events over the NC background events, which rarely contain an extended track-like structure.

The neutrino energy is measured as the sum of the muon energy and the hadronic shower energy. The muon energy is measured using the range of the muon in the case that it is fully contained within the detector, and using curvature in the magnetic field if it exits the detector. The hadronic shower energy is measured using a kNN that looks at aspects of the shower profile to return the energy. This method was found to give an improved energy resolution over using a pure calorimetric energy measurement (as used in the first two MINOS analyses), reducing the energy resolution from 55\% to 43\% for 1.0-1.5$\,$GeV showers, for example~\cite{kNNShwThesis}.  

The selected fiducial events are binned as a function of reconstructed neutrino energy, and those CC \numu{} events in the FHC beam are further divided by their estimated energy resolution to improve sensitivity~\cite{kNNShwThesis,jmThesis,sjcThesis}. For the RAF events, only the muon energy is considered and hence the events are binned in reconstructed muon energy only.


The predicted FD CC \numu{} or CC \antinumu{} energy spectrum is calculated using a combination of simulation and the ND data. The beam flux simulation and the method used to tune it based on the ND data was described in Section \ref{sec:beamFluxSim}. The transport of the simulated particles through the detector simulation is performed by the GCALOR~\cite{gcalor} and GEANT3~\cite{geant3} packages. The process to calculate the FD prediction, known as the extrapolation procedure, consists of the following steps. Firstly, the event selection is performed at the ND for both data and simulation. The simulation is used to produce a matrix that converts between the reconstructed and true neutrino energy. The selected ND data are then multiplied by this matrix to convert to a pseudo-true energy. At this stage a correction is also applied to account for the selection efficiency in the ND. The next step applies the beam matrix, a correction that accounts for the difference in acceptance of the neutrino beam between the two detectors (the beam appears as a point source for the FD, whereas the ND sees an extended source). With the energy spectrum now in pseudo-true energy, the neutrino oscillations are applied using the three flavour (or historically, two flavour) oscillation formalism. Finally, the FD selection efficiency is applied and the energy is converted back to reconstructed neutrino energy using the FD version of the reconstructed to true energy conversion matrix. The resultant energy spectrum from this process is the predicted spectrum for the FD for the given oscillation parameters used in the extrapolation procedure. 

\subsection{Atmospheric Neutrinos}\label{sec:atmosDis}
The atmospheric neutrino selection aims to select those neutrinos produced in cosmic ray interactions in the upper atmosphere. These interactions give rise to both muon and electron type neutrinos from the decay of pions and muons, as shown previously for the production of the neutrino beam in Eqs \ref{eq:piDecay} and \ref{eq:muDecay}. The total exposure to atmospheric neutrinos over the lifetime of the MINOS experiment amounted to 37.88$\,$kt-yrs.

The atmospheric neutrino sample is collected exclusively by the FD since the location deep in the Soudan Mine gives a large reduction in the background events coming from cosmic rays. The atmospheric neutrino interactions are then separated from the remaining cosmic background by looking for events that have their interaction vertex inside the fiducial volume (contained-vertex sample), or by looking for muon-like, upward-going events entering the detector from the bottom region of the detector (non-fiducial muons)~\cite{chThesis,abThesis,jdcThesis}. The cosmic ray background in the contained-vertex sample is further reduced by checking for activity in the cosmic ray veto shield associated with the main detector event.

The contained-vertex and non-fiducial muon selections are divided into candidate CC \numu{} and CC \antinumu{} samples depending on the measured charge of the muon in the event. In the two-flavour MINOS analyses, these data were binned as a function of $\log_{10}(L/E)$ but the binning scheme was changed for the three flavour analysis. In the three flavour analysis, the data are binned in two dimensions as a function of $\log_{10}(E)$ and the zenith angle $\cos\theta_z$. This scheme was chosen in order to maximise the sensitivity to the MSW effect, and hence the mass hierarchy, that modifies the oscillation probability as the neutrinos travel through the interior of the earth, where the distance travelled depends on the measured value of $\cos\theta_z$.

Lastly, a selection is made to identify contained-vertex shower events. This selection consists mainly of NC $\nu$, CC \nue{} and CC \antinue{} interactions. These events have limited sensitivity to neutrino oscillations but are all included in the fit in a single bin to constrain the absolute atmospheric neutrino flux~\cite{speakmanThesis}. 

The simulation of atmospheric neutrinos is based on the Bartol flux predictions~\cite{bartolFlux}. Atmospheric neutrinos that interact inside the fiducial volume, the contained-vertex sample, are simulated using NEUGEN3~\cite{neugen} in the same way as for the beam neutrinos. NUANCE~\cite{nuance} is used to simulate the interaction of the atmospheric neutrinos in the rock surrounding the cavern and to propagate the final state particles up to the edge of the detector. The simulation of the particles in the detector is then the same for both samples, using the GCALOR~\cite{gcalor} and GEANT3~\cite{geant3} packages, in exactly the same way as for the beam neutrino simulation.
Efforts are made to account for the change in atmospheric neutrino fluxes as a function of time due to variations in the solar cycle, an important consideration since the period over which data were collected covers nearly an entire solar cycle. It is predicted that the atmospheric neutrino flux can vary by up to 7\% over this period~\cite{minosAtmos2012}.

Oscillations are applied to the FD predicted energy spectra using a reweighting technique for all of the data samples listed in Sections \ref{sec:beamDis} and \ref{sec:atmosDis}.  The process includes the addition of the backgrounds from \nutau{} and \antinutau{} appearance. The oscillations applied were historically those derived from the two-flavour approximation, but in the final MINOS analysis described in this article the full three flavour formalism was used. The oscillation parameters are then varied during the fit in order to extract the parameters that provide the best fit to the data.

\subsection{Two Flavour Oscillation Results}
Oscillations are applied to the FD predicted energy spectra using a reweighting technique for all of the data samples listed in Sections \ref{sec:beamDis} and \ref{sec:atmosDis}.  The process includes the addition of the backgrounds from \nutau{} and \antinutau{} appearance. The early analyses performed by MINOS, as well as other experiments, considered the oscillations in terms of an approximate two neutrino case. In the limit that \thetareac{} tends to zero, all but one of the additional terms shown in Eq. \ref{eq:3flavDis} that modified the two-flavour approximation vanish, leaving just $\Delta m^2 = \Delta m^2_{32} + \sin^{2}\theta_{12}\Delta m^2_{21}$. Since $\sin^{2}\theta_{12}\Delta m^2_{21} / \Delta m^2_{32} \approx 0.01$ then \dmsol{} could easily be ignored when measurements of $\Delta m^{2}$ were much less accurate than 1\%. Thus the sector governing \numu{}$\rightarrow$\numu{} oscillations could be considered as decoupled from the solar scale oscillations.

The first measurement from MINOS of $\Delta m^2$ and $\sin^{2}2\theta$ was made in 2006 using 1.27$\times$10$^{20}\,$POT~\cite{minosPRL2006} and was followed by updated analyses using exposures of 3.36$\times$10$^{20}\,$POT~\cite{minosDis2008} and 7.25$\times$10$^{20}\,$POT~\cite{minosDis2011}.  The first measurement of the antineutrino oscillation parameters $\Delta \overline{m}^2$ and $\sin^{2}2\overline{\theta}$  was made in 2008 using 1.71$\times$10$^{20}\,$POT~\cite{minosAntiDis2011} and was followed by a further analysis using an exposure of 2.95$\times$10$^{20}\,$POT~\cite{minosAntiDis2012}.  Measurements of both the neutrino and antineutrino oscillation parameters were also made using just the atmospheric neutrino oscillation sample, exploiting the complete MINOS atmospheric neutrino sample of 37.88$\,$kt-years~\cite{minosAtmos2012}. 

\begin{figure}
  \centering
  \includegraphics[scale=0.38]{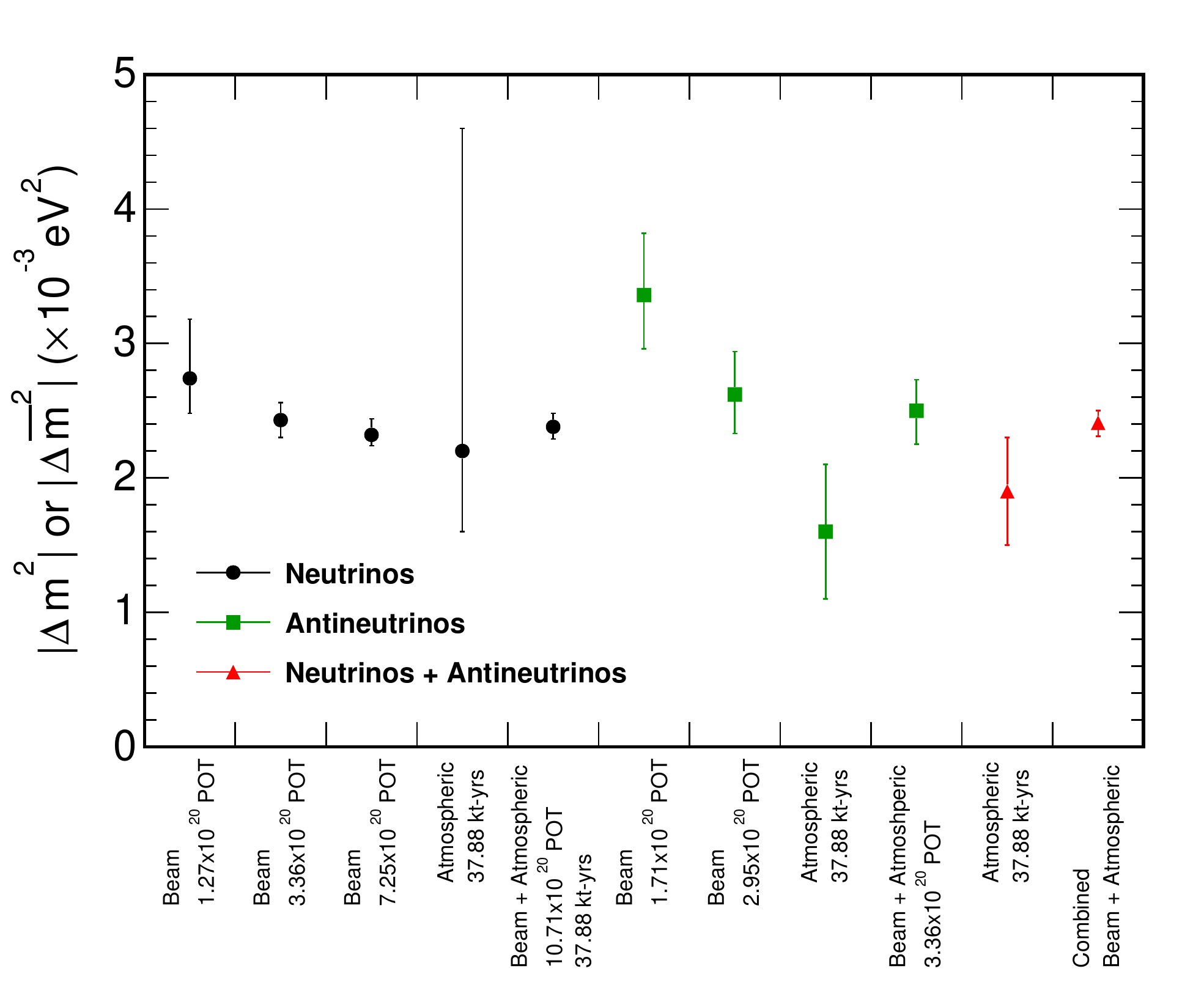}
  \caption{\label{fig:minosHistory}The values of $|\Delta m^{2}|$ and $|\Delta\overline{m}^{2}|$ as measured by MINOS using the two-flavour approximation throughout the lifetime of the experiment ~\cite{minosPRL2006,minosDis2008,minosDis2011,minosNuNubar,minosAntiDis2011,minosAntiDis2012,minosAtmos2012}. The black points show measurements made using CC \numu{} interactions, and those in green using CC \antinumu{} interactions. The two red points show the combination of CC \numu{} and CC \antinumu{} events under the assumption that the oscillation parameters are identical between neutrinos and antineutrinos. The $x$-axis provides details of the beam and atmospheric neutrino exposure used to produce the measurement.}
\end{figure}

\begin{figure}
\centering
\includegraphics[scale=0.4]{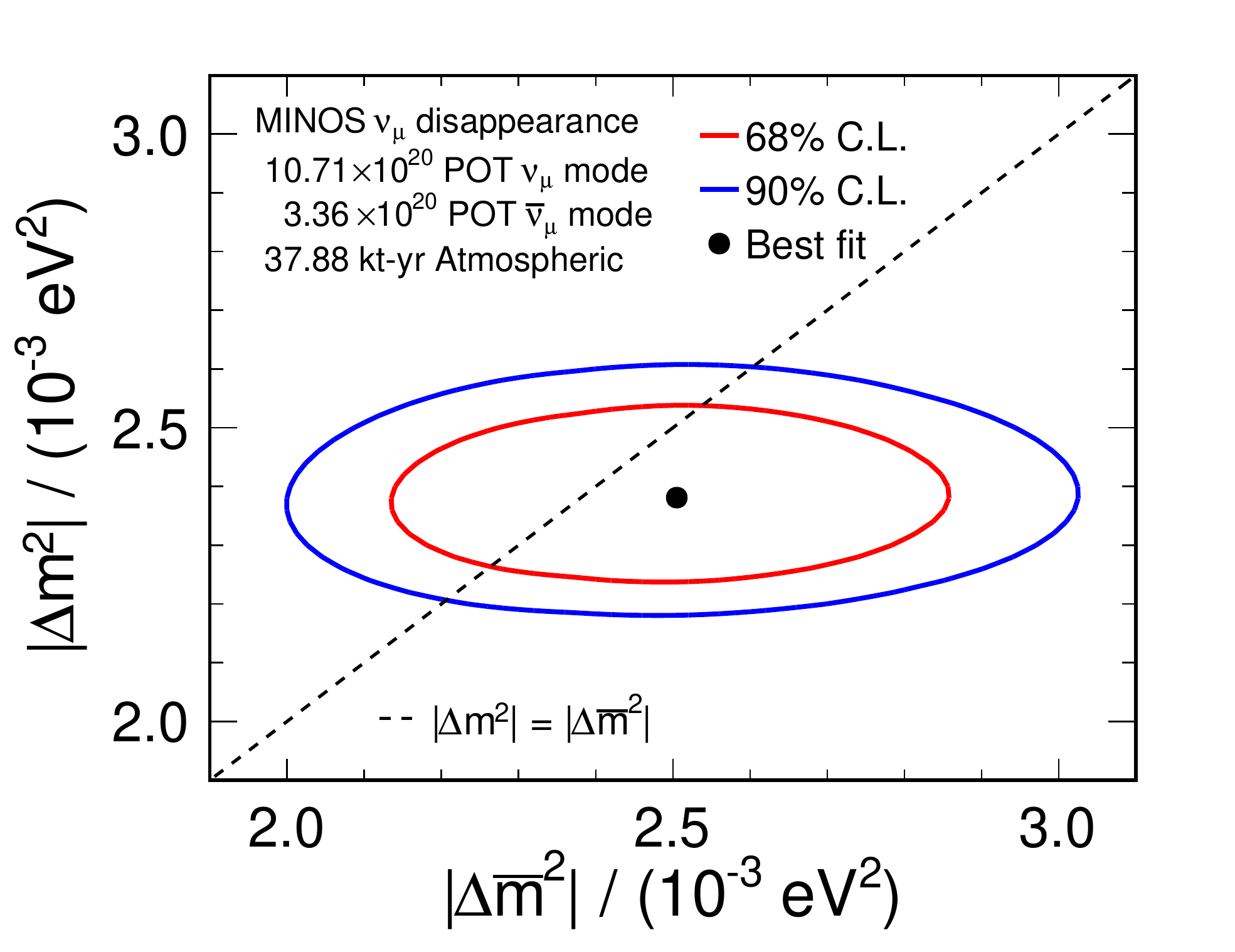}
\caption{\label{fig:dmdmbar}A comparison of the MINOS measurements of the two-flavour oscillation mass-splitting governing long-baseline \numu{} and \antinumu{} disappearance, $|\Delta m^{2}|$ and $|\Delta\overline{m}^{2}|$, showing both the 68\% (red) and 90\% (blue) confidence limit contours~\cite{minosNuNubar}. }
\end{figure}

The final two-flavour fit~\cite{minosNuNubar} considered the full 10.71$\times$10$^{20}\,$POT in FHC mode, 3.36$\times$10$^{20}\,$POT in RHC mode and the 37.88$\,$kt-years of atmospheric neutrinos. This fit was performed both using four parameters (meaning that different oscillation parameters were used to fit the neutrinos and antineutrinos) and two parameters (where neutrinos and antineutrinos are assumed to oscillate in the same way). The values and 1$\sigma$ uncertainties of $\Delta m^2$ measured from these analyses are summarised in Fig. \ref{fig:minosHistory}, showing good agreement between measured values for neutrinos and anti-neutrinos with the full exposure. Figure \ref{fig:dmdmbar} explicitly shows the agreement between the values of $\Delta m^2$ and $\Delta\overline{m}^{2}$ measured in the four parameter version of the final fit, and hence that the parameters that govern oscillations of neutrinos and antineutrinos are the same within the uncertainty of the measurement, allowing all of the samples to be considered together to fit the parameters of the PMNS matrix in the full three flavour fit described in Section \ref{sec:3flavResults}. The values of $\Delta m^2$ and $\Delta\overline{m}^{2}$ were the most accurate measurements made of the two flavour mass-splitting, but the values of $\sin^{2}2\theta$ and $\sin^{2}2\overline{\theta}$ had a larger uncertainty compared to those measured by Super-K~\cite{superKPRL2011}.
 
\section{Electron Neutrino Appearance}
MINOS is also able to search for the subdominant appearance of electron neutrinos in the muon neutrino beam. This channel, being subdominant, must always been considered in the case of three neutrino flavours. The main aim of the search is to perform a measurement of \thetareac{}, with the main measurable being given by the first term in Eq. \ref{eq:3flavApp} as $\sin^{2}\theta_{23}\sin^{2}2\theta_{13}$.

The preselection of candidate events begins with the requirement that the events must occur in time with the neutrino beam. Additionally, the events must be consistent in direction with the beam. Shower-like topology events are then selected by requiring that the event does not have a track-like object of at least 25 planes, or extending at least 15 planes from the edge of the shower. The events are also required to have at least five consecutive planes with energy deposits of at least one half of the energy deposit expected from a minimum ionising particle. Only the region in energy where the majority of \nue{} and \antinue{} appearance is expected is used in the analysis, limiting the allowed reconstructed neutrino energy to be within the range from 1 to 8$\,$GeV.

The candidate CC \nue{} and CC \antinue{} interactions are then identified using the library-event-matching (LEM) method whereby each data event is compared on a hit-by-hit basis to a vast library of 20 million simulated signal (CC \nue{}  for FHC beam data or CC \antinue{} for RHC beam data) and 30 million background (NC) events~\cite{jporThesis,rbtThesis,apsThesis}. As discussed in Section \ref{sec:minosInt}, it is not possible separate CC \nue{} and CC \antinue{} interactions in this analysis. The best 50 matches to the data event are used to calculate a series of variables that are combined to form a single PID variable, called $\alpha_{LEM}$, using an artificial neural network. All those events with values of $\alpha_{LEM}$ above 0.6 are selected as part of the analysis, a cut value defined to maximise the sensitivity to the \nue{} and \antinue{} appearance signal.

Selected events are binned in two dimensions as a function of reconstructed neutrino energy and $\alpha_{LEM}$. The bins with values of $\alpha_{LEM}$ closer to one have the most sensitivity to oscillations since they have the highest purity of CC \nue{} and CC \antinue{} events.

\subsection{FD Prediction}
Due to the absence of \nue{} and \antinue{} appearance in the ND, different methods are used to predict the expected background and signal components of the FD energy spectrum.

The three main backgrounds to the appearance signal come from NC interactions, CC \numu{} or CC \antinumu{} events, and intrinsic beam CC \nue{} and CC \antinue{} interactions. These three backgrounds are measured using the ND using the same binning scheme used in the main event selection for both data and simulation~\cite{joaoThesis}. A selection is then performed using simulation at the FD, and for each bin in energy and $\alpha_{LEM}$ the bin content for each background component is multiplied by a correction factor from the ND equal to the ratio of the number of data to simulation events. The small \nutau{} and \antinutau{} appearance background must be calculated in a different way since, like the CC \nue{} and CC \antinue{} appearance, it does not occur in the ND. This background is derived from the simulation and then corrected using the ND measurement of CC \numu{} or CC \antinumu{} events.

The ND can not be used directly to measure the signal efficiency due to the lack of an appearance signal at such a short baseline. Instead, a sample of CC \numu{} interactions are selected from data. The energy deposits in these interactions arising from the muon are then removed from the event~\cite{amhThesis} and replaced with energy deposits from a simulated electron shower~\cite{jaabThesis}. The simulated electron vertex, direction and energy are set to match those of the reconstructed muon in the original data event. This procedure makes an effective sample of CC \nue{} and CC \antinue{} data events that can be used to study the efficiency of selecting and identifying the signal events. This method was validated using those events that are not sensitive to the appearance signal, defined by $\alpha_{LEM} < 0.5$, to predict the number of events in the same region of the FD data and agreement was found within the 0.3(0.6)$\sigma$ of the statistical uncertainty for the CC \nue{}(\antinue) sample~\cite{minosApp2012}.

\subsection{Results}
\begin{figure}
\centering
\includegraphics[scale=0.35]{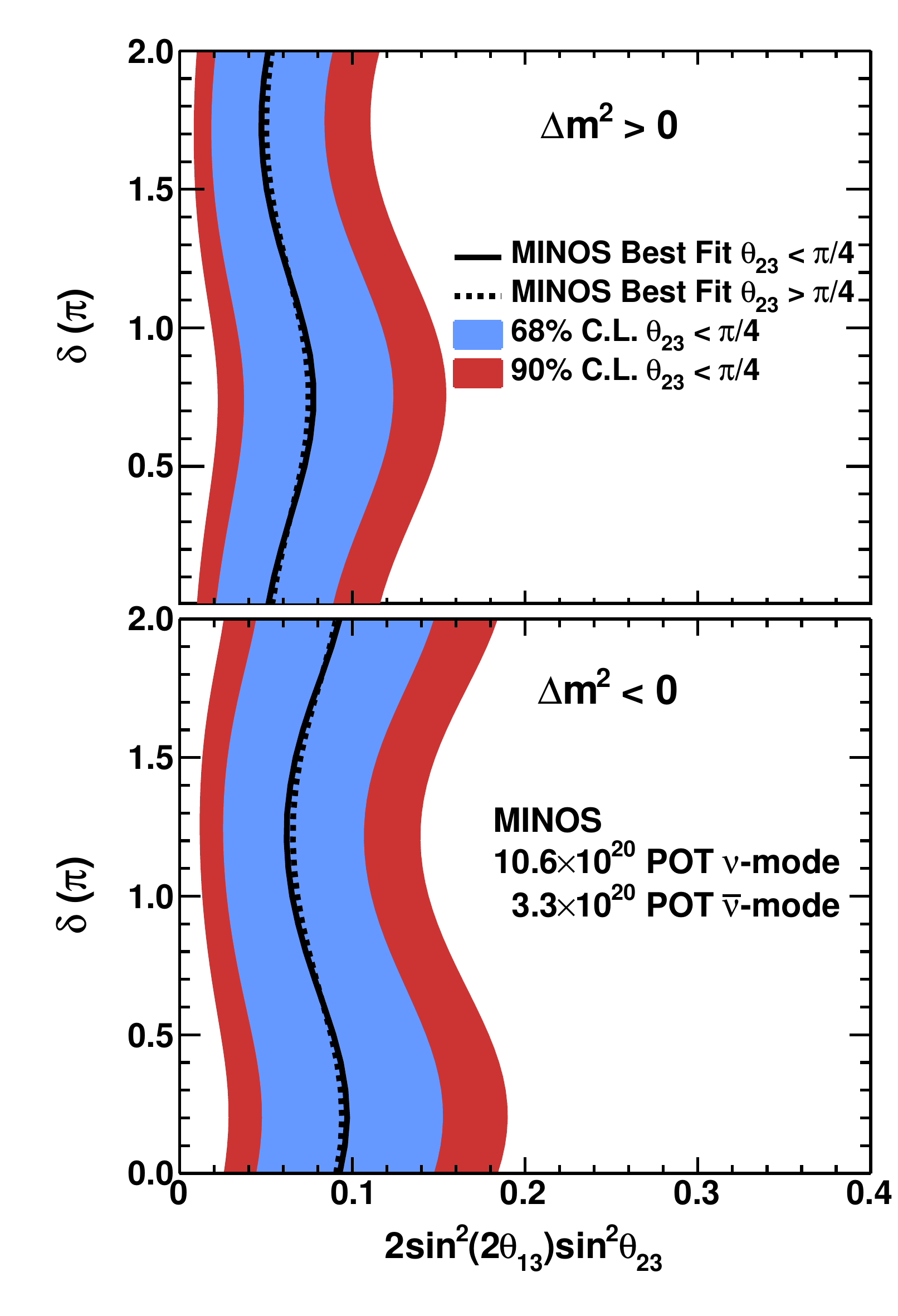}
\includegraphics[scale=0.4]{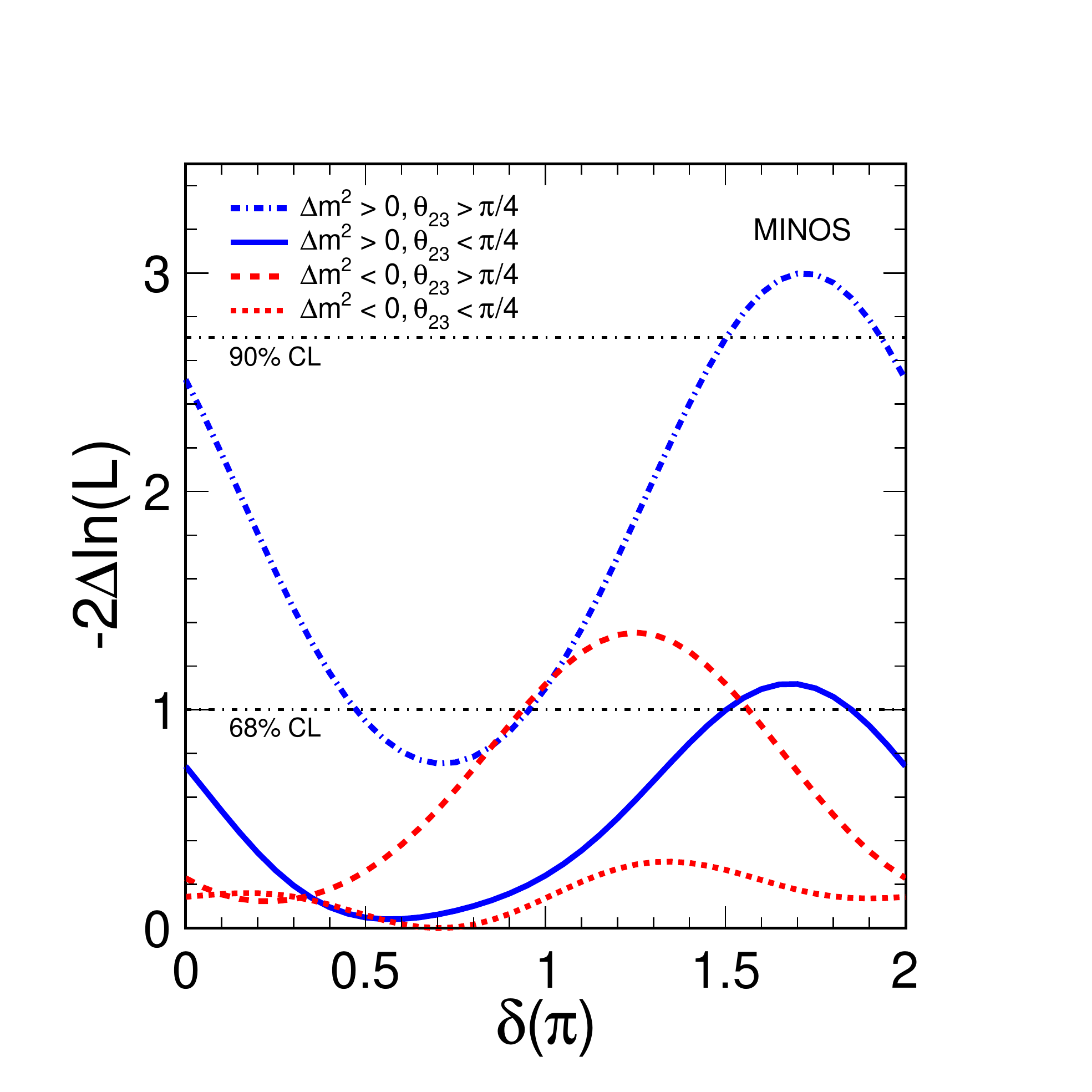}
\caption{\label{fig:nueAppContour}Left: the \numu{}$\rightarrow$\nue{} appearance contour shown as a function of $2\sin^{2}2\theta_{13}\sin^{2}\theta_{23}$ and \deltacp{}. The normal hierarchy is shown in the top panel, and the inverted hierarchy below, with the 68\% and 90\% contours shown for the lower octant of \thetaatm{}. The best fit curve is also shown for the upper octant, showing little sensitivity to the octant of \thetaatm{}. Right: the likelihood shown as a function of \deltacp{} for the four combinations of mass hierarchy and \thetaatm{} octant. Those likelihood values above the horizontal lines are disfavoured at the 68\% and 90\% C.L. Figure from Ref.~\cite{minosApp2012}.}
\end{figure}

MINOS has performed two \numu{}$\rightarrow$\nue{} searches on data samples of 7.01$\times$10$^{20}\,$POT \cite{minosApp2010} and 8.20$\times$10$^{20}\,$POT \cite{minosApp2011}, and a combined \numu{}$\rightarrow$\nue{} and \antinumu{}$\rightarrow$\antinue{} search based on a total exposure of 10.6$\times$10$^{20}\,$POT FHC and 3.3$\times$10$^{20}\,$POT RHC~\cite{minosApp2012}. The POT of the final analysis does not agree exactly with those quoted for the muon neutrino disappearance analysis because the short high energy run is not included in the electron neutrino appearance analysis as it has no sensitivity to the appearance signal. The result of the combined \nue{} and \antinue{} appearance search is shown on the left of Fig. \ref{fig:nueAppContour}, excluding the null hypothesis of no appearance at approximately the 96\% confidence level. The contours are shown for the normal (top) and inverted (bottom) hierarchy for the lower octant of \thetaatm{}. The best fit is also shown for the upper octant, showing little sensitivity to the octant of \thetaatm{}. This analysis found the value of \thetareac{} to be greater than zero with less significance than the T2K result from 2011~\cite{t2kPRL2011} and the reactor experiments from 2012~\cite{dayaBay2012,reno2012,doubleChooz2012}.

The data were also used to study the mass hierarchy, value of \deltacp{} and the octant of \thetaatm{} by using an external constraint from the reactor neutrino experiments $\sin^{2} 2\theta_{13} = 0.098 \pm 0.013$~\cite{dayaBay2012,reno2012,doubleChooz2012}. The likelihood is shown as a function of \deltacp{} for the four combinations of mass hierarchy and \thetaatm{} octant on the right of Fig. \ref{fig:nueAppContour}. The sensitivity to these parameters is low, but this represents the first attempt from a long-baseline neutrino oscillation experiment to constrain these parameters and lays the foundation for future measurements.

\section{Combined Three Flavour Analysis}\label{sec:3flavResults}

\begin{figure*}
\centering
\includegraphics[scale=0.7]{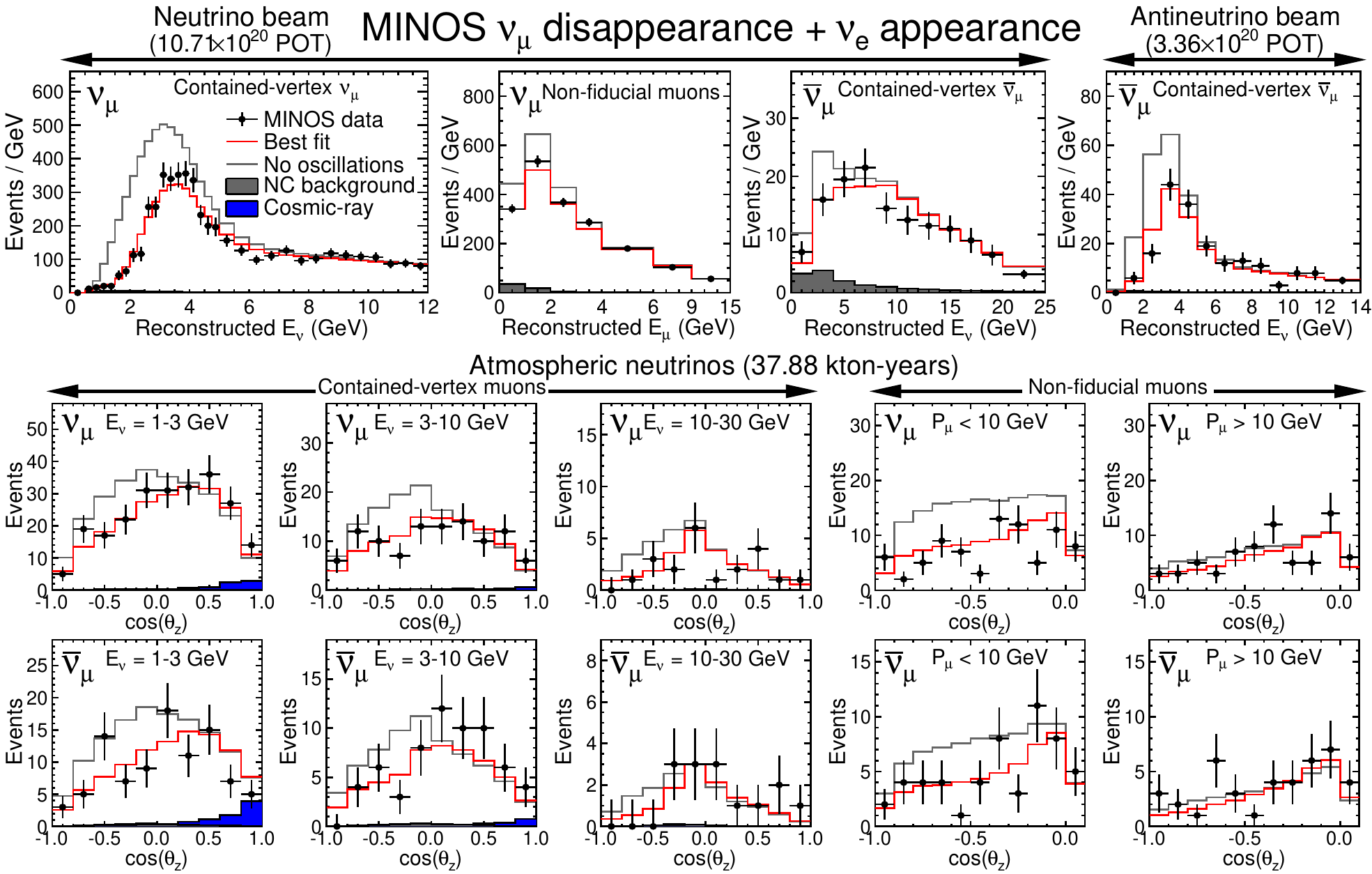}
\caption{\label{fig:eventSpec}The event spectra from 10.71$\times$10$^{20}\,$POT FHC mode, 3.36$\times$10$^{20}\,$POT RHC mode and 37.88$\,$kt-yrs of atmospheric data. The data are shown compared to the prediction in absence of oscillations (grey line) and to the best-fit prediction (red). The beam histograms (top) also include the NC background component (filled grey) and the atmospheric histograms (bottom) include the cosmic-ray background contribution (filled blue).}
\end{figure*}

The analysis outlined here uses the full three flavour oscillation framework to perform a combined fit of the beam and atmospheric CC \numu{} and CC \antinumu{} disappearance samples along with the CC \nue{} and CC \antinue{} appearance samples. The CC \numu{} and CC \antinumu{} event spectra from this analysis are shown in Fig. \ref{fig:eventSpec}. All data are shown compared to both the null oscillations prediction (grey) and the best fit prediction with oscillations (red). The beam data (top row) also show the background from NC events (filled grey) and the atmospheric data is also compared to the background arising from cosmic-ray muons (filled blue).

\begin{figure}
\centering
\includegraphics[scale=0.35]{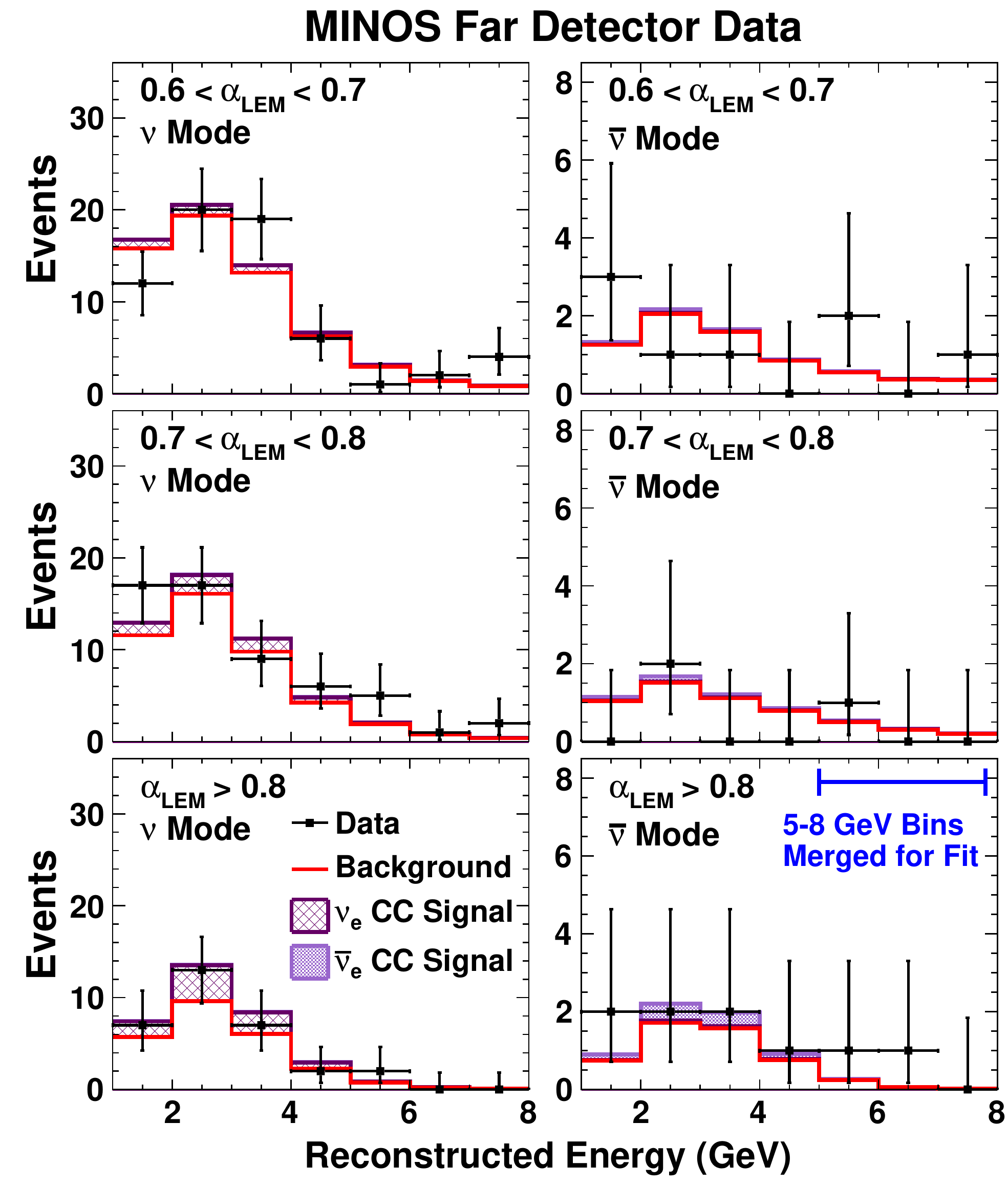}
\caption{\label{fig:nueSpec}The \nue{} (left) and \antinue{} (right) candidate energy spectra, each shown in three bins of $\alpha_{LEM}$, the PID parameter. The data in black are shown compared to the expected background (red line) and the expected three flavour appearance signal (purple for CC \nue{} and lilac for CC \antinue{}). The bins from 5-8$\,$GeV are combined for all samples in the fit. }
\end{figure}

The \nue{} and \antinue{} appearance data are shown in Fig. \ref{fig:nueSpec}, binned as a function of reconstructed energy and $\alpha_{LEM}$. The bins between 5 and 8$\,$GeV are shown for display purposes, but are combined into a single bin in the fitting procedure.

A combined fit of all the MINOS data allows for the maximum extraction of information on the mass hierarchy, octant of \thetaatm{} and the value of \deltacp{}. The appearance sample is sensitive to \deltacp{}, and it also provides some information on the mass hierarchy via matter effects as well as a small sensitivity to the octant of \thetaatm{}. The atmospheric neutrino sample includes a resonance region in multi-GeV upward going events, again providing sensitivity to the mass-hierarchy. This resonance exists in the normal hierarchy for neutrinos and in the inverted hierarchy for antineutrinos.

The oscillation parameters that are free within the fit are \dmatm{}, $\sin^{2}\theta_{23}$, $\sin^{2}\theta_{13}$ and \deltacp{}. The value of $\sin^{2}\theta_{13}$ is constrained using a Gaussian penalty term in the fit, using the central value and 1$\sigma$ uncertainty from the average of the results from the reactor neutrino experiments Daya Bay~\cite{dayaBay2013}, RENO~\cite{reno2012} and Double Chooz~\cite{doubleChooz2012}: $\sin^{2}\theta_{13}  = 0.0242 \pm 0.0025$. The values of \dmsol{} and $\sin^{2}\theta_{12}$ are kept fixed in the fit at the following values: $\Delta m^{2}_{21}=7.54\times10^{-5}\,$eV$^2$ and $\sin^{2}\theta_{12}= 0.307$~\cite{fogli}. The effect of varying \dmsol{} and $\sin^{2}\theta_{12}$ was checked by shifting them by their quoted uncertainty and looking at the change in the fitted values of  \dmatm{} and $\sin^{2}\theta_{23}$. The changes caused by varying these parameters were found to be negligible, hence no penalty terms are included for \dmsol{} and $\sin^{2}\theta_{12}$.

The oscillation probabilities used to perform the fit are calculated directly from the PMNS matrix without assumptions. The method takes advantage of matrix manipulation algorithms specially designed for high computational speed~\cite{fastPMNS}. Matter effects are included using a four layer approximation of the PREM model~\cite{PREM1981}. All of the systematic uncertainty parameters are included with the corresponding samples and treated as nuisance parameters with penalty terms in the fit. The systematic parameters are those that account for the main differences between the simulation and the data. The likelihood is calculated separately for the \numu{} disappearance and \nue{} appearance samples and the two contributing values are then summed together under the assumption that the systematic uncertainties in the two samples are uncorrelated.  

\begin{figure}
  \centering
  \includegraphics[scale=0.50]{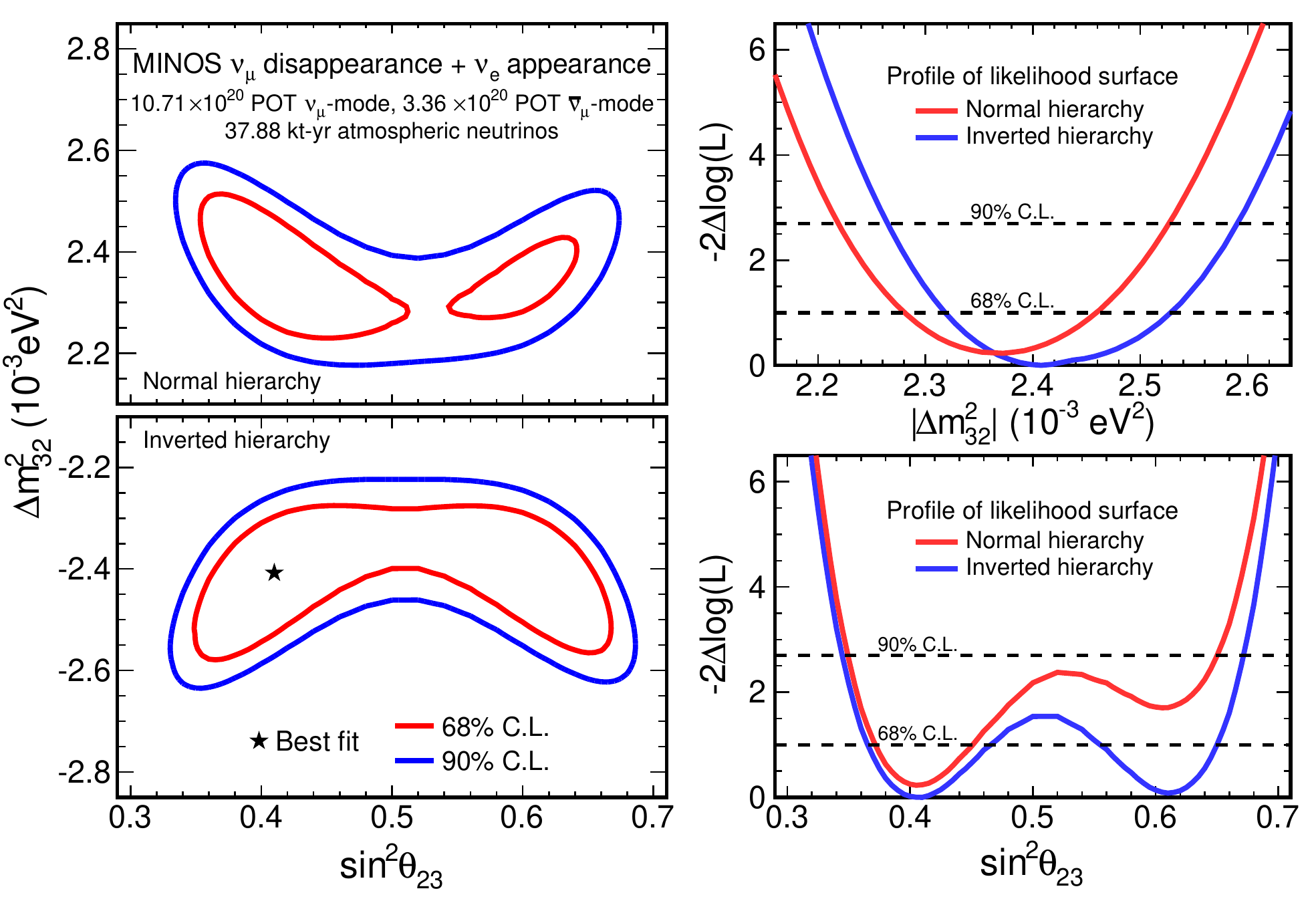}
  \caption{\label{fig:minosCombi}Left: The 68\% (red) and 90\% (blue) confidence limit contours as a function of \dmatm{} and $\sin^{2}\theta_{23}$ from the combined analysis of 10.71$\times$10$^{20}\,$POT FHC data, 3.36$\times$10$^{20}\,$POT RHC data and 37.88$\,$kt-years of atmospheric neutrinos. Right: The profiled 1D likelihood for \dmatm{} (top) and $\sin^{2}\theta_{23}$ (bottom) assuming both normal hierarchy (red) and inverted hierarchy (blue). The best fit point lies in the inverted hierarchy, lower octant quadrant, and there is a slight tendency to disfavour the normal hierarchy, upper octant region.}
\end{figure}

The 2D confidence limits for \dmatm{} and $\sin^{2}$\thetaatm{}, calculated by maximising the log-likelihood at each point in the 2D parameter space with respect to $\sin^{2}$\thetareac{}, \deltacp{} and all of the systematic parameters, is shown in Fig. \ref{fig:minosCombi}. The 68\% contour is shown in red and the 90\% contour is shown in blue. The overall best fit point is found to be in the inverted hierarchy, lower octant region. The results are  $|$\dmatm{}$|=[2.28 - 2.46]\times10^{-3}\,$eV$^{2}$ (68\%) and $\sin^{2}$\thetaatm{}${}=0.35-0.65$ (90\%) in the normal hierarchy $|$\dmatm{}$|=[2.32 - 2.53]\times10^{-3}\,$eV$^{2}$ (68\%) and $\sin^{2}$\thetaatm{}${}=0.34-0.67$ (90\%) in the inverted hierarchy. These measurements of $|$\dmatm{}$|$ are the most precise at the time of writing, but Super-K~\cite{superKPRD2010} and T2K~\cite{t2kPRL2014} have higher accuracy measurements of \thetaatm{}. The case known as maximal mixing in two flavour oscillations, namely that \thetaatm{}${}=\pi/4$, is disfavoured at 1.54 units of $-2\Delta\log(\mathcal{L})$~\cite{minosCombined}. 

\begin{figure}
  \centering
  \includegraphics[scale=0.38]{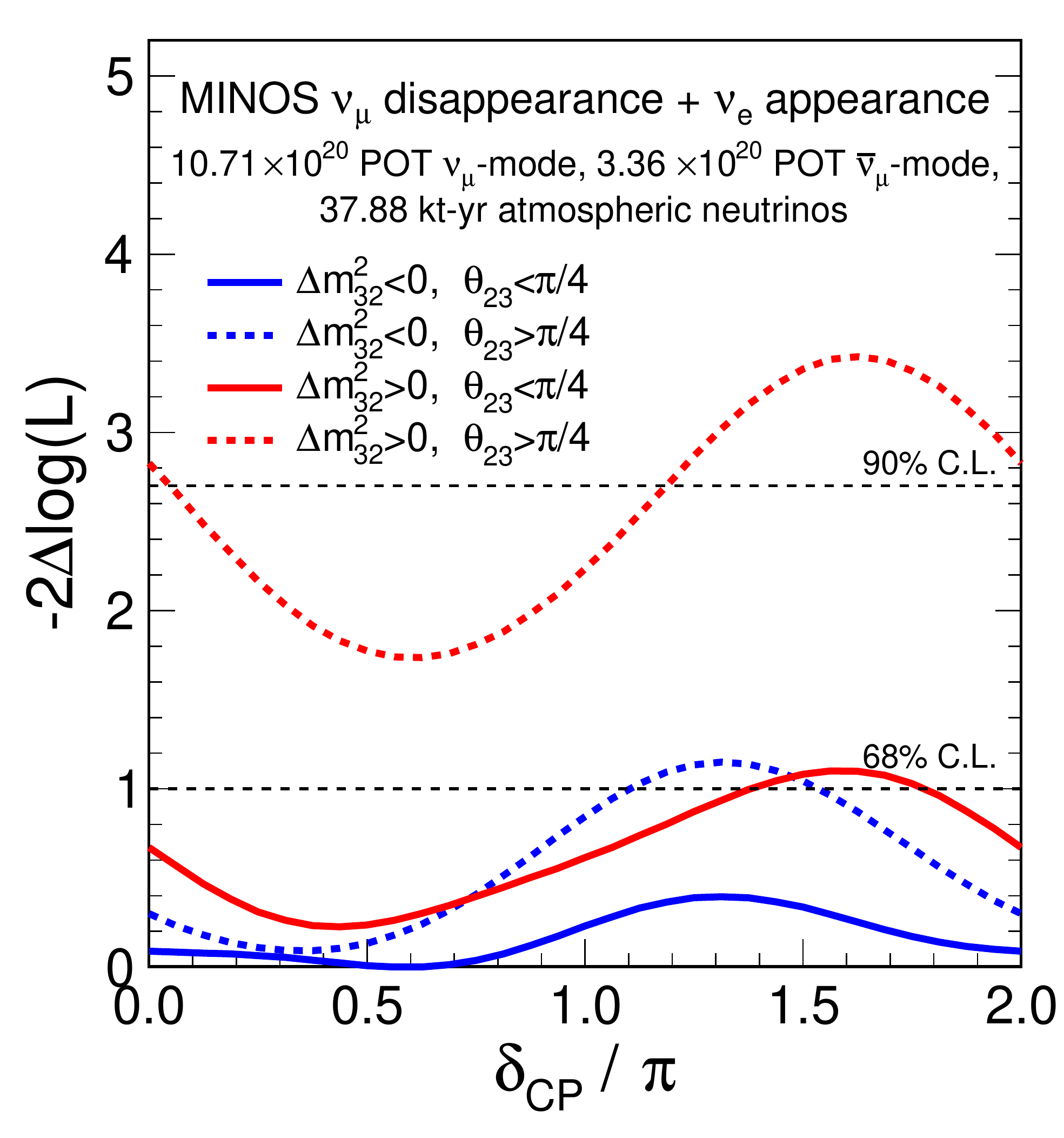}
  \caption{\label{fig:minosCP}The likelihood surface profiled as a function of $\delta_{CP}$ under the four choices of mass hierarchy and $\sin^{2}\theta_{23}$ octant. The normal hierarchy upper octant case is disfavoured at 90\% for half of the range of \deltacp{}.}
\end{figure}

Figure \ref{fig:minosCP} shows the 1D likelihood profile as a function of \deltacp{}. This distribution shows an enhanced sensitivity compared to the \nue{}+\antinue{} appearance only result shown in the right plot in Fig. \ref{fig:nueAppContour}, but less than that of T2K~\cite{t2kPRL2014NuE}. The best-fit oscillation parameters are shown in Table \ref{tab:fitResults} for each combination of the mass hierarchy and octant of \thetaatm{}. The data slightly disfavour the normal hierarchy, upper octant case across the whole range of \deltacp{}, and above 90\% for approximately half of the range of \deltacp{}, with the best fit point in that case being disfavoured by a $-2\Delta\log(\mathcal{L})$ of 1.74. The other three choices of the mass hierarchy and octant have very similar values of $-2\Delta\log(\mathcal{L})$ and remain degenerate.

\begin{table*}
  \centering
  \begin{tabular*}{0.9\textwidth}{ccccccc}
  \hline \hline \\[-2ex]
  \vspace{2pt}
  Mass Hierarchy & \thetaatm{} Octant & \dmatm{}${}/10^{-3}\,$eV$^{2}$ & $\sin^{2}\theta_{23}$ & $\sin^{2}\theta_{13}$ & \deltacp{}${}/\pi$ &  $-2\Delta\log(\mathcal{L})$\\
  \hline \\[-2ex]
  \dmatm${}<0{}$ & \thetaatm${}<\pi/4$ & -2.41 & 0.41 & 0.0243 & 0.62 & 0 \\
  \dmatm${}<0{}$ & \thetaatm${}>\pi/4$ & -2.41 & 0.61 & 0.0241 & 0.37 & 0.09 \\
  \dmatm${}>0{}$ & \thetaatm${}<\pi/4$ & 2.37 & 0.41 & 0.0242 & 0.44 & 0.23 \\
  \dmatm${}>0{}$ & \thetaatm${}>\pi/4$ & 2.35 & 0.61 & 0.0238 & 0.62 & 1.74 \\
  \\[-2.0ex]
  \hline \hline
  \end{tabular*}
  \caption{\label{tab:fitResults}The fit results from the combined disappearance and appearance analysis. The best-fit values of the oscillation parameters for the four combinations of the mass hierarchy and octant of \thetaatm{}. Also shown is the difference in $-2\Delta\log(\mathcal{L})$ calculated relative to the overall best-fit point that is found in the inverted hierarchy, lower octant region. This table was reproduced from Ref.~\cite{minosCombined}.}
\end{table*}

\section{Three Flavour Oscillations with MINOS+}
The neutrino energy spectrum from the ME tune of the NuMI beam for MINOS+ is shown by the dashed line in Fig. \ref{fig:fluxConfig}, compared to MINOS LE configuration (solid) and the high energy configuration (dotted), and peaks between 3$\,$GeV to 10$\,$GeV. This means that MINOS+ probes the oscillation paradigm in the tail of the neutrino oscillation spectrum. It is in this region that more exotic phenomena such as sterile neutrinos or large extra dimensions are more easily seen from the distortion of the oscillation signal. These searches will not be discussed here, but the high statistics measurement of oscillations away from the oscillation maximum provide a stringent test of three flavour neutrino oscillations.

MINOS+ collected a total of 2.99$\times$10$^{20}\,$ POT in the first year of running from September 2013 until September 2014. The ND reconstruction software was rewritten in order to better cope with the higher rate of interactions produced by the upgraded NuMI beam, both in terms of minimising the impact of event pile-up and by increasing the speed of the algorithms to facilitate the prompt processing of data. A main focus of this effort was to prevent tracking failures, which provided a considerably improved ND efficiency.

\begin{figure}
\centering
\includegraphics[scale=0.38]{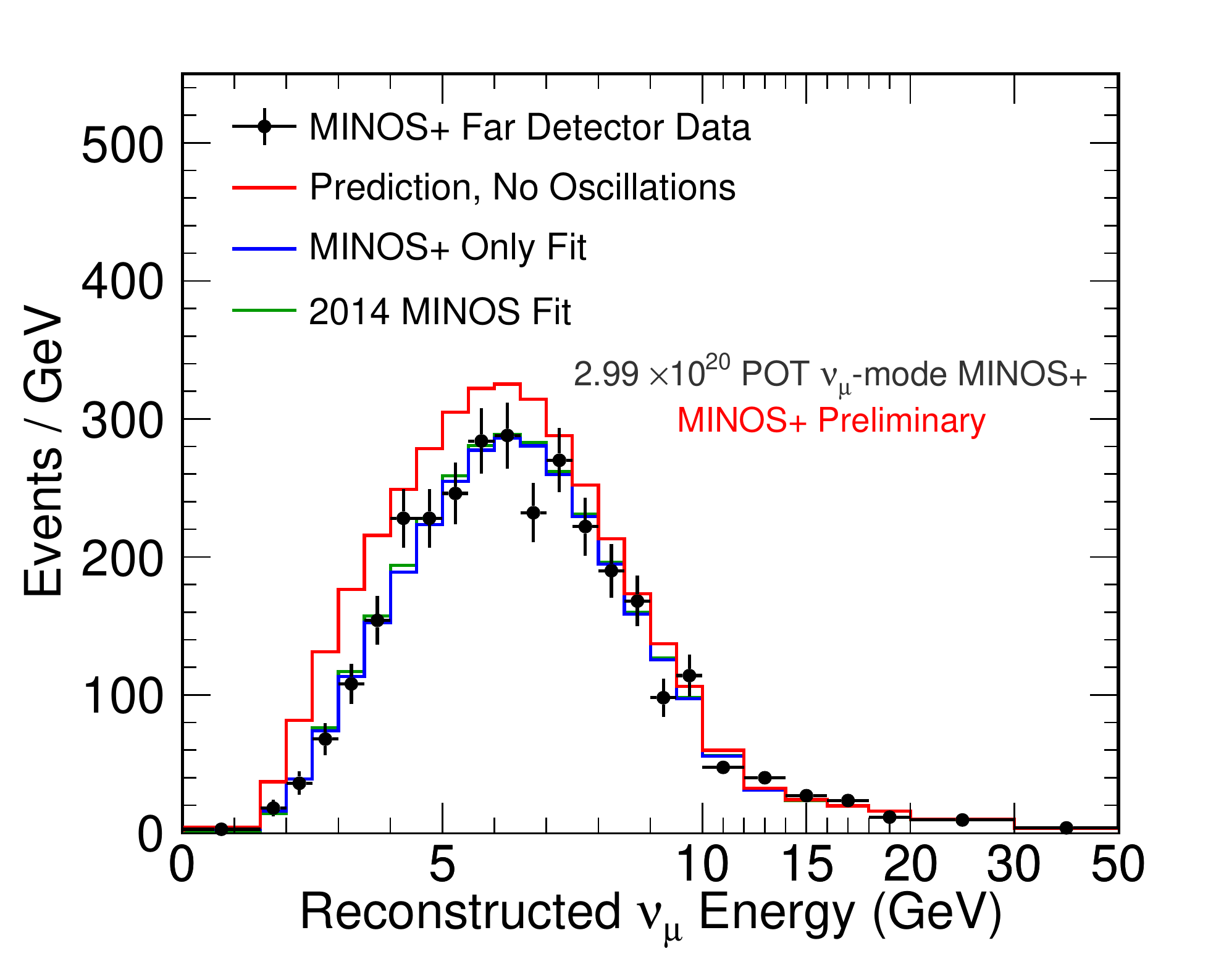}
\includegraphics[scale=0.38]{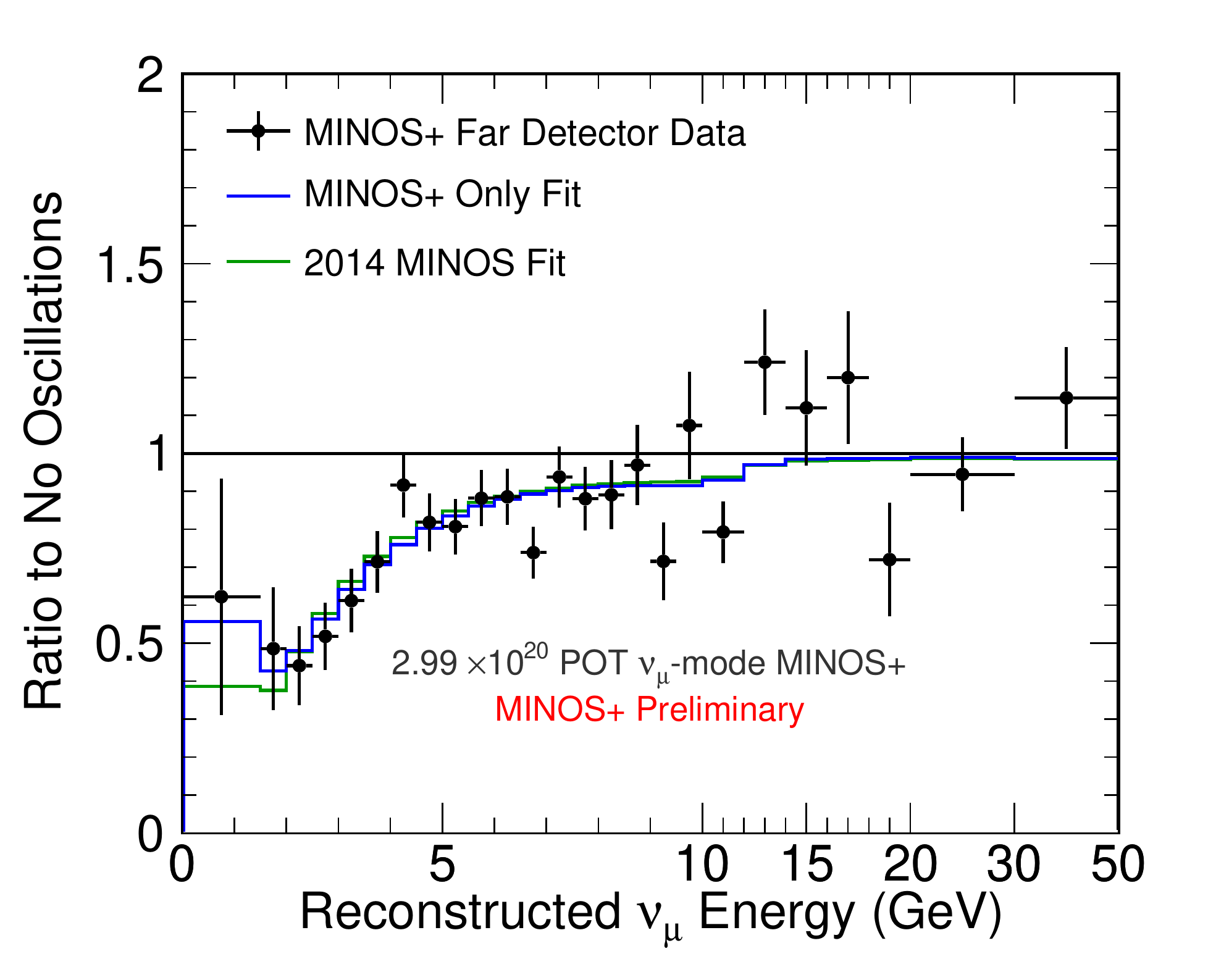} 
\caption{\label{fig:minosPlusData}The reconstructed CC \numu{} energy spectrum for beam neutrinos for MINOS+ (left) and the ratio of data to the unoscillated MC (right) for an exposure of 2.99$\times10^{20}\,$POT. The red curve shows the prediction in the case of no oscillations, the blue curve shows the best fit to the MINOS+ data alone, and the green curve shows the combined fit result from MINOS.}
\end{figure}

In the first instance, a fit to just the MINOS+ data sample was performed such that the best fit oscillation parameters could be compared to those measured by MINOS. The reconstructed CC \numu{} energy spectrum is shown in the left plot of Fig. \ref{fig:minosPlusData} and is also shown as a ratio to the no oscillations case on the right. The unoscillated prediction is shown in red along with two oscillated predictions: the blue shows the MINOS+ only best fit and the green shows the best fit using the parameters measured by the final combined MINOS analysis, as described in Section \ref{sec:3flavResults}. The blue and green curves are very consistent, showing that the oscillation parameters measured by MINOS clearly provide a good description of the MINOS+ data. 

\begin{figure}
\centering
\includegraphics[scale=0.38]{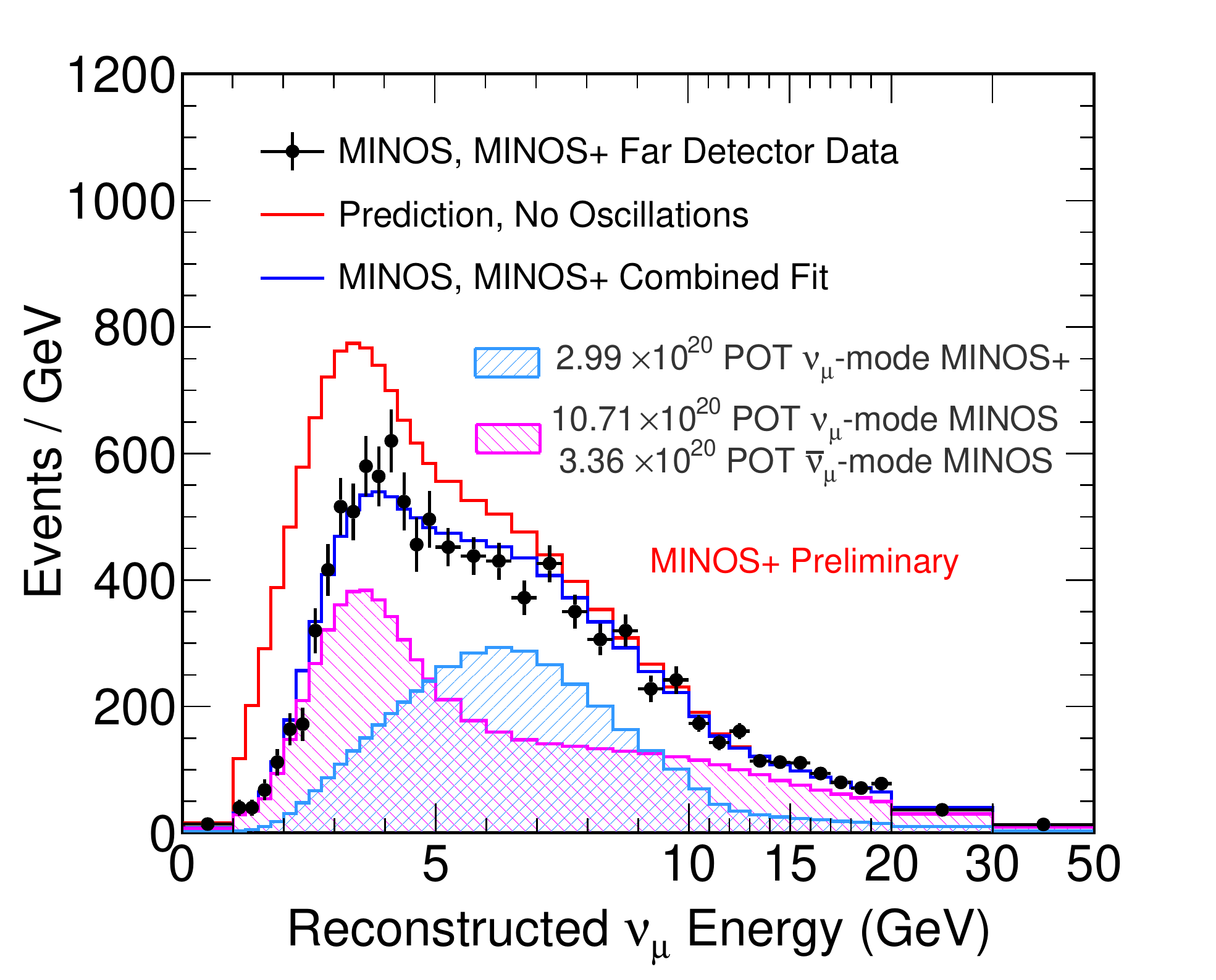}
\caption{\label{fig:minosMinosPlusSpec}The reconstructed CC \numu{} energy spectrum for beam neutrinos in MINOS and MINOS+. The red curve shows the prediction in the case of no oscillations, and the blue curve shows the best fit prediction. The two filled histograms show the components of the best fit corresponding to MINOS (pink) and MINOS+ (blue).}
\end{figure}

A combined fit of all the data included in the MINOS combined analysis and the 2.99$\times10^{20}\,$POT of MINOS+ data was also performed. The combined reconstructed CC \numu{} energy spectrum is shown in Fig. \ref{fig:minosMinosPlusSpec} compared to the best fit prediction from oscillations in blue. The MINOS and MINOS+ components of the best fit prediction are shown in the pink and blue filled histograms, respectively, showing that the statistics in the region around $6-8\,$GeV have doubled with only about a third of the expected exposure for MINOS+. The difference in the best fit point between the final MINOS result and this combined fit was $-2\Delta\log(\mathcal{L})=1.3$. The 2D contours in \dmatm{} and $\sin^{2}\theta_{23}$ are not shown as only a small improvement is seen compared to the combined MINOS analysis. This is expected since the MINOS+ energy distribution only provides a fairly small sensitivity to the oscillation parameters compared to MINOS. The result will be updated with the data from the full MINOS+ exposure.

\section{Conclusion}
The MINOS experiment collected data from the NuMI beam over a period spanning 2005 until 2012, and atmospheric neutrinos at the FD from 2003 until 2012, putting it at the forefront of neutrino oscillation physics for a decade. The combined analysis of the CC \numu{} and CC \antinumu{} disappearance samples, coming from both beam and atmospheric neutrino sources, and the CC \nue{} and CC \antinue{} appearance samples using a full three flavour fit produced the world's best measurement of the atmospheric mass splitting \dmatm{}. Whilst the sensitivity to the mass hierarchy, octant of \thetaatm{} and the \emph{CP}-violating phase \deltacp{} is small, the data tend to disfavour the combination of normal mass hierarchy and upper octant of \thetaatm{} at the 90\% confidence level across half of the \deltacp{} phase-space. The overall best fit point was measured to be in the inverted hierarchy, lower octant region.

The first year of data from the MINOS+ experiment, using the ME beam configuration, was analysed and shown to give very consistent results compared to the values of the neutrino oscillation parameters measured in the final MINOS analysis. Further data collected by MINOS+ will provide future stringent tests of the standard neutrino oscillation paradigm and allow for investigations of more exotic phenomena such as sterile neutrinos, non-standard interactions and large extra dimensions.

\section{Acknowledgements}
The work of the MINOS and MINOS+ collaborations is supported by the US DOE, the UK STFC, the US NSF, the State and University of Minnesota, the University of Athens in Greece, and Brazil's FAPESP and CNPq. We are grateful to the Minnesota Department of Natural Resources, the crew of the Soudan Underground Laboratory, and the personnel of Fermilab, for their vital contributions.

\section{Bibliography}
\bibliography{minosReview}

\end{document}